\documentclass[12pt]{article}                     
\begin{document}

\noindent{\bf Nuclear Resonant scattering of Synchrotron radiation
from  nuclei in the Brownian motion}
\vskip 1.0 cm
\noindent{\bf Ashok Razdan}
\vskip 1.5 cm
\noindent Nuclear Research Laboratory,\\ 
Bhabha Atomic Research Centre,\\ 
Trombay, Mumbai-400 085\\
\vskip 1.0 cm
\noindent PACS No: 76.20.+q,76.80.+y,05.40.+jc,07.85.+Qe
\vskip 1.0 cm
\noindent Keywords:Resonant Scattering, Synchrotron radiation, Brownian motion.
\vskip 1.0 cm
\noindent{\bf Abstract:}

The time evolution of  the coherent forward scattering of 
the synchrotron radiation for resonant nuclei in  Brownian motion
is studied. Apart from target thickness, the appearance of the dynamical
beats also depends on
'$\alpha$' which is the ratio of the harmonic
force constant to the damping force constant of  harmonic
oscillator undergoing Brownian motion.

\vfill\eject

\noindent{\bf Introduction:}

In recent years , synchrotron radiation (SR) has been used to
study M\"ossbauer
spectroscopy in the time domain. By measuring the time dependence of the
intensity of nuclear de-excitation [1 and references therein] various
solid state and nuclear
parameters are being studied. Synchrotron radiation is produced
when ultra-relativistic
electrons or positrons  accelerate  while transgressing
bending magnets or periodic magnetic structures
and emit intense radiation. SR is emitted in short repetitive pulses
whereas emission of photons from a radioactive nucleus are
random in time.

The M\"ossbauer spectra of iron-protein crystals have special
shape  which provide a unique signature,
indicating that M\"ossbauer atom undergoes bounded diffusive motions in the
biological systems. It is observed  for biological systems
that above a particular temperature,
which is the characteristic of the system (i)the M\"ossbauer
lineshape becomes
non-Lorentzian, (ii) the mean square displacement (msd) shows temperature
dependence which is not typical of either  the Einstein or the  Debye model
and (iii) the width of the M\"ossbauer lineshape increases with the
temperature. The non-Lorentzian lineshape was observed
by  Parak et. al. [2], Cohen et. al.[3]
in  myoglobin. Similar phenomena was
seen in haemoglobin by Mayo,Parak and M\"ossbauer [4].
Pioneering studies on the dynamics of proteins using M\"ossbauer Spectroscopy
were done by Keller and Debrunner [5], Parak et. al. [6,7].
These unusual results observed for biological systems,
have been explained by using a dynamical model of the bounded diffusion.
These models are based on the concept of Brownian motion of an overdamped
harmonic oscillator [2,3,8,9]. Recently
synchrotron radiation has been used to study dynamics of M\"ossbauer atom in
the biological systems [10,11,12,13].
The goal of the present paper is to study the time evolution of
resonant scattering of SR from nuclei undergoing
Brownian motion in harmonic potential.

\noindent{\bf Methodology:}

In this study we use  mathematical treatment developed in 
reference [8,9] as a starting point. A short pulse of SR can
be decomposed into a continuum set of monochromatic spectral components
within a frequency interval. Smirnov and Kohn [14,15,20]  calculated the forward
transmitted wave packet by integrating over all spectral components of the
transmitted radiation to obtain 
\begin{equation}
E(t,Z)=\frac{E_{\omega 0}(Z)}{2\pi }\int d\omega \exp (-i\omega t)\exp(\frac{iK C(k,\omega )Z}{2})
\end{equation}

E(t,Z) is the time dependent electric field amplitude of the synchrotron
radiation transmitted through a nuclear target of thickness Z. The function 
$E_{\omega 0} (Z)$ has the modulus $(\frac{I_0}{\Delta\omega})^{1/2}$ $\exp(-\frac{
\mu_eZ}{2})$ . The electronic absorption
coefficient is given as $\mu_e$=K $\chi$
where $\chi$ =$\frac{\sigma_e}{V_0 K}$,  $\sigma_e$ being the
total cross section of electronic inelastic scattering and
$V_0$ is the target volume corresponding to one nucleus.
K is the wave number defined as 
K=$\frac{\omega}{c}$ and $\omega$ is the frequency having wave vector k. 
$I_0$ is the intensity of the SR within the frequency band $\Delta\omega$,
as determined by the monochromator system.
$C(k,\omega)$ is the nuclear part of the susceptibility of the target and is
related to the scattering amplitude and can be written as
\begin{equation}
C(k,\omega)=\frac{i\Gamma_{0}}{2\hbar}\sum_{ge} B_{ge} \phi(k,\omega-\omega_{eg})
\end{equation}
where
\begin{equation}
B_{ge}=\frac{8 \pi f_{LM}(k)}{\omega^{2} V_{o}(2I_{g}+1)\Gamma_{0}} |<g|j^{~}(k)|e>|^{2}
\end{equation}
characterizes the strength of the nuclear response at the resonant frequency
$\omega_{eg}$ and
$\phi(k,\omega-\omega_{eg})$ is the universal resonance function.
Here,g and e define sublevels of the ground and the excited states
of nucleus respectively and $ \Gamma_{0}$ is the natural linewidth of the excited levels.
$f_{LM}(k)$ is the Lamb M\"ossbauer factor, $I_g$ is the  nuclear spin in
the ground state while $<g|j^{~}(k)|e>$ represents the matrix element of the
scaler component of the nuclear current density vector.

At resonance , nuclear part of the susceptibility of the target has a frequency
dependence , which is determined by the the nature of universal resonance
function, obtained (in the present case) by taking
into account the diffusive motion. It is described by
the following
\begin{equation}
\phi(k,\omega) =\int_{0}^{\infty} dt\exp(-i\omega t-\frac{\Gamma_{0}t}{2\hbar}
)F_s(k,t)
\end{equation}
where                  
\begin{equation}
F_s(k,t)=\int dr\exp(-ik.r) G(r,t)
\end{equation}

Here G(r,t) is the correlation function. The derivation of above equations
is given in references [14,15] in detail. We consider only that case where
polarization is absent in the coherent forward scattering. The most frequent
application of SR in the nuclear resonance spectroscopy is the  measurement of
the time-dependence of the forward scattering intensity, given by
\begin{equation}
I_f(t,z)=|E(t,Z)|^{2}
\end{equation}
It is interesting to note that M\"ossbauer absorption spectrum
is dependent on the real part of the universal resonance
function $\phi(k,\omega)$. On the other hand, the time dependence is
determined by the entire universal function.

\noindent{\bf Harmonically bound nuclei in Brownian motion:}

The time evolution of nuclear resonant scattering
of SR pulse by nuclei undergoing
various types of diffusion (like free diffusion, continuous
localised and jump diffusion)  has been
studied earlier [14,15] . Here we consider an important case of
diffusion in the form of
Brownian motion. 
Gunther et. al. [16,17]
were the first to predict that Brownian motion of a M\"ossbauer
atom results in
the non-Lorentzian nature of M\"ossbauer line shape. The theory
for the cases
of motion of harmonically bound M\"ossbauer atom in the Brownian motion has
been extensively developed both for the classical [2,3,8,9]
and quantum cases [18].
In the present paper we will confine to the classical case and use the
theory as developed by Nowik et. al. [8,9]. For one-dimensional
harmonic oscillator
in the Brownian motion, Uhlenbeck and Ornstein [19] derived a general formula
for self correlation function $G(x,x_{0},t)$ which is the probability that
at time t the nucleus will be at position $x$ if at time t=0 it was at
position $x_{0}$. This self correlation is given as [8] 
\begin{equation}
G(x,x_{0},t)=(\frac{\alpha }{2{\pi \ D(1-exp(-2\alpha t)}})^{1/2}\exp (\frac{
-\alpha (x-x_{0}exp(-\alpha t))^{2}}{2D(1-exp(-2\alpha t))})
\end{equation}

where 
\begin{equation}
\alpha =\frac{w^{2}}{\beta }
\end{equation}
is the ratio of the harmonic force constant $w$
and damping force constant $\beta$.
The diffusion constant D is given by
\begin{equation}
D=\frac{{k_{B}}T}{m\beta }
\end{equation}
Nowik et. al.[8,9] generalised this self-correlation
function to three dimensions for the over-damped case
and obtained a simple formula for the line shape.
M\"ossbauer spectra were computed for  range of $\alpha$ values
and fixed diffusion constant D.
The parameter $\alpha$ [8,9] decides
the nature of the M\"ossbauer lineshape. 
The Mossbauer line-shape is non-Lorentzian for smaller values of
$\alpha$ [8,9]. Since M\"ossbauer lineshape for
biological systems is non-Lorentzian in nature , we will focus on
smaller values of $\alpha$ in our calculations.

The universal resonance function
characteristic of the Brownian motion can be written as [9] 
\begin{equation}
\phi(k,\omega )=\exp (\frac{-k^{2}D}{\alpha })\sum_{m=0}\frac{1}{m!}(\frac{k^{2}D}{
\alpha })^{m}\frac{i {t _0}}{(\omega +i t_0+im {\alpha })}
\end{equation}
where 
\begin{equation}
t_{0}=\frac{\Gamma _{0}}{2\hbar }
\end{equation}
Thus,in the present case, for equation (1),  $C(k,\omega )$ is given by
\begin{equation}
C(k,\omega )=\sum_{ge}\sum_{m}\frac{B_{ge} t_{0}a_{k}(m)}{(\omega -\omega
_{0})+i(t_{0}+m\alpha )}
\end{equation}
where
\begin{equation}
a_{k}(m)=\frac{1}{m!}(\frac{k^{2}D}{\alpha})^{m}\ exp(-\frac{k^{2}D}{\alpha })
\end{equation}
Thus E(Z,t) is given as
\begin{equation}
E(Z,t)=\frac{E_{\omega _{0}}(Z)}{2\pi }\int d\omega \ exp(-i\omega t)\exp {(-
\frac{{\mu _{n}}Z{t_{0}}\sum_{m}{a_{k}}(m)}{(\omega -\omega
_{0})+i(t_{0}+m\alpha )}})
\end{equation}
where 
\begin{equation}
\mu _{n}=K\sum_{ge} B_{ge}
\end{equation}
The summation over 'ge' in equation (15) is related to the summing over the
transition between the ground and the excited states of nucleus [ 14,15 ].
It is important to note that for any particular value of $(\frac{k^{2} D}{\alpha})$
,only a few Lorentzians contribute to the universal line shape [9].

\noindent{\bf Results:}

To illustrate the role of
Brownian motion in the time dependence of forward scattering,we first
consider a case  of a thin target (single line sample)
having an effective  resonance thickness $\mu _{n}$ Z=1. The results
of the numerical calculation for various values of $\alpha $
are shown in the Figure 1. It is evident from this figure, that
 the natural decay appears
(manifested by straight segment of time response curve) 
faster for smaller values of $\alpha $ and slower for larger values
of $\alpha $. The time response in the
case of  single-line target, having an effective resonant thickness $\mu
_{n}$ Z=10 (thick samples) for various values of $\alpha $,
are shown in the Figure 2  where
dynamical beats appear.It is also clear from this
figure that smaller the
value of $\alpha $, earlier the dynamical beat occurs and vice versa.
The appearance of the
dynamical beats is also dependent on the value of $\mu_n$ Z.
Comparing Figures
1 and 2 it is clear that $\alpha $ and $\mu_{n}$ Z share a definite
relationship,with each other so for as the appearance of dynamical beats are
concerned. In order to further investigate this relationship we fix the value
of $\alpha $ and see how the variation of $\mu _{n}$ Z affects the appearance
of dynamical beats in the time response behaviour. The results are shown in
the Figure 3 which indicates that larger the value of $\mu _{n}$ Z,
earlier does the dynamical beat appear and vice versa.

\noindent{\bf Comparison with diffusion results:}

In general the M\"ossbauer lineshape has a Lorentzian character.
The diffusion of the M\"ossbauer atom causes broadening
in this Lorentzian lineshape. For the case of free diffusion, the increase
in the M\"ossbauer linewidth is proportional to  the diffusion constant.
The presence of the this diffusion constant in the exponential factor
of the time response 
causes an accelerated decay  of the coherent
signal.  
However, the position of the
appearance of the 'dynamical
beat minimum ' is  independent of the value of the
diffusion constant. All dynamical beats appear at
the same time with variable minimum values for different values of the
diffusion constants. But in the cases of bounded diffusion 
inside a potential well and  jump diffusion, the universal
resonance functions have  complicated shapes represented, in general,
by the coherent superposition of the Lorentzian lineshapes. The width and
weight of each Lorentzian is determined by the specific nature
of the diffusion process. Thus, the appearance of the dynamical beats in the
time response is sensitive to the diffusion rate and cage-size for
the cases of the jump diffusion and bounded diffusion,respectively.

The universal resonance function for
the case of Brownian motion in general (and for smaller values of $\alpha$
in particular) is non-Lorentzian.This is because the universal resonance
function is the result of  superposition of various Lorentzian line shapes.
Each Lorentzian line shape
has its characteristic weight and width as determined by the value
of $\alpha$ . This type of the nature of universal resonance function
results in the
complicated behaviour in the time response. As in the cases of continuous
diffusion and jump diffusion, the physical reason for the complicated
behaviour in the time response for the case of Brownian motion,is the split
of the universal resonance function into several terms. 

\noindent{\bf Conclusion:}
                                   
The method of coherent forward scattering
of the SR reveals complex amplitudes
of the oscillation of the electromagnetic field in the presence of the
Brownian motion. 
\vfill\eject
\noindent{\bf Figure captions:}

Figure 1:  The time dependence of nuclear forward scattering of the 
synchrotron radiation in the presence of Brownian motion for different values
for a thin sample of effective thickness $\mu_n$ Z=1 corresponding
to  $\alpha$ = 0.5, 0.6, 0.7, 0.8, 0.9, 1.0, 2.0, 4.0 .
The lowermost curve corresponds to $\alpha$=0.5 and uppermost
curve corresponds to $\alpha$=4.0.
$\alpha$ is units of $sec^{-1}$.

Figure 2:  The time dependence of nuclear forward scattering of the 
synchrotron radiation in the presence of Brownian motion for different values
of $\alpha$ for a thick sample of effective thickness $\mu_n$ Z=10.
$\alpha$ is units of $sec^{-1}$.
(1) $\alpha$ = 0.5;(2) $\alpha$=0.6;(3) $\alpha$=0.7; (4) $\alpha$=0.8;
(5) $\alpha$ =0.9; (6) $\alpha$ =1.0; (7) $\alpha$ =2.0 ; (8) $\alpha$=4.0 .

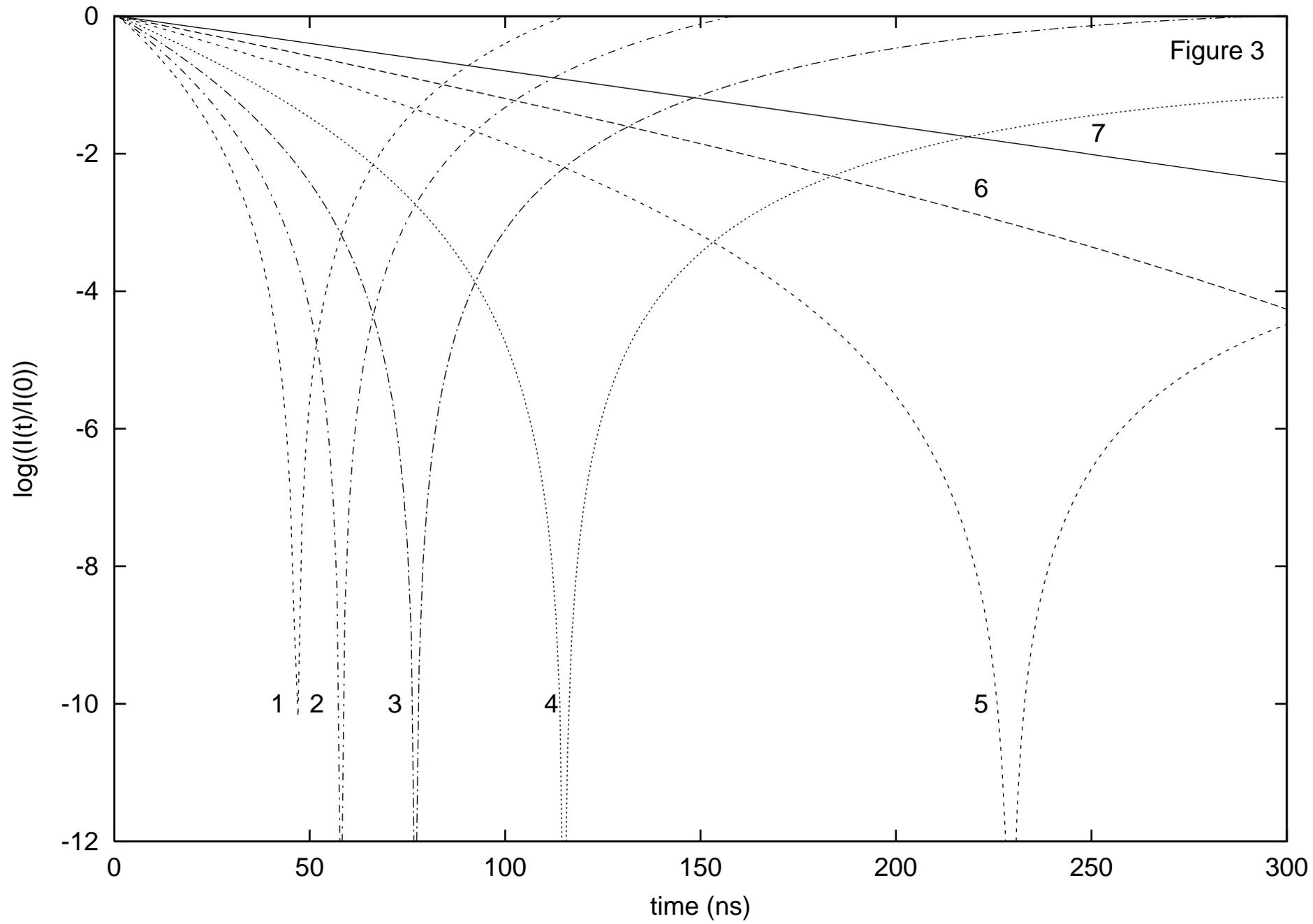
Figure 3:  The time dependence of nuclear forward scattering of the 
of synchrotron radiation  with $\alpha$=1 for different values of
$\mu_n$ Z. $\alpha$ is in units of $sec^{-1}$
(1) $\mu_n$ Z = 50.0;(2) $\mu_n$ Z=40.0;(3) $\mu_n$ Z=30.0;(4) $\mu_n$ Z=20.0;
(5) $\mu_n$ Z =10.0; (6) $\mu_n$ Z =5.0 ; (7) $\mu_n$ Z =1.0 .

\vfill\eject
\noindent{\bf References:}
\begin{enumerate}
\item W.Sturhahn, E.E. Alp, T.S.Toellner, P.Hession ,M.Hu and J.Sutter
      Hyperfine Interactions 113(1998)47
\item F.Parak,E.N.Frolov, R.L.Mossbauer and V.I.Goldanskii, J of Mol.
      Biology,145(1981)825
\item S.G.Cohen, E.R.Bauminger, I.Nowik, S.Ofer, J.Yariv
      Phys.Rev.Lett.  46(1981)1244
\item K.H.Mayo, F.Parak,R.L.Mossbauer, Phys.Lett. A 82(1981)468
\item H.Keller and P.G.Debrunner, Phys.Rev.Lett.,45(1980)68
\item F.Parak,E.W.Knapp,D.Kucheida, J. Mol. Biol. 161(1982)177
\item F.Parak et. al. Hyperfine Interactions,58(1990)2381
\item I.Nowik, E.R.Bauminger, S.G. Cohen, S.Ofer,Phys.Rev.Lett. 50(1983)1528
\item I.Nowik, E.R.Bauminger, S.G.Cohen, S.Ofer, Phys.Rev.A 31(1985) 2291
\item A.X.Trantwein and H.Winkler, Hyperfine Interactions 123/124(1999)561
\item H.Grunsteudel et. al. Inorganica Chimica Acta 275-276(1997)1334
\item C.Keppler et. al. European Biophys. J . 25(1997)221
\item F.Parak and K.Achterhold, Hyperfine Interactions 123/124(1999)825
\item G.V.Smirnov and V.G.Kohn,Phys. Rev. B 52(1995)3356
\item G.V.Smirnov and V.G.Kohn,Phys. Rev. B 57(1998)5788
\item L.Gunther, J.Zitkova-Wilcox, J.Phys. (France) 35(1974)c6-519
\item L.Gunther, J.Zitkova-Wilcox, J.Stat.Phys. 12(1975)205
\item A.Razdan, Eur.Phys.J.B 8(1999)143
\item G.E.Uhlenbeck, L.S.Ornstein,Phys. Rev. 36(1930)823
\item V.G.Kohn,G.V.Smirnov, Hyperfine Interactions 123(1999/2000)327
\end{enumerate}
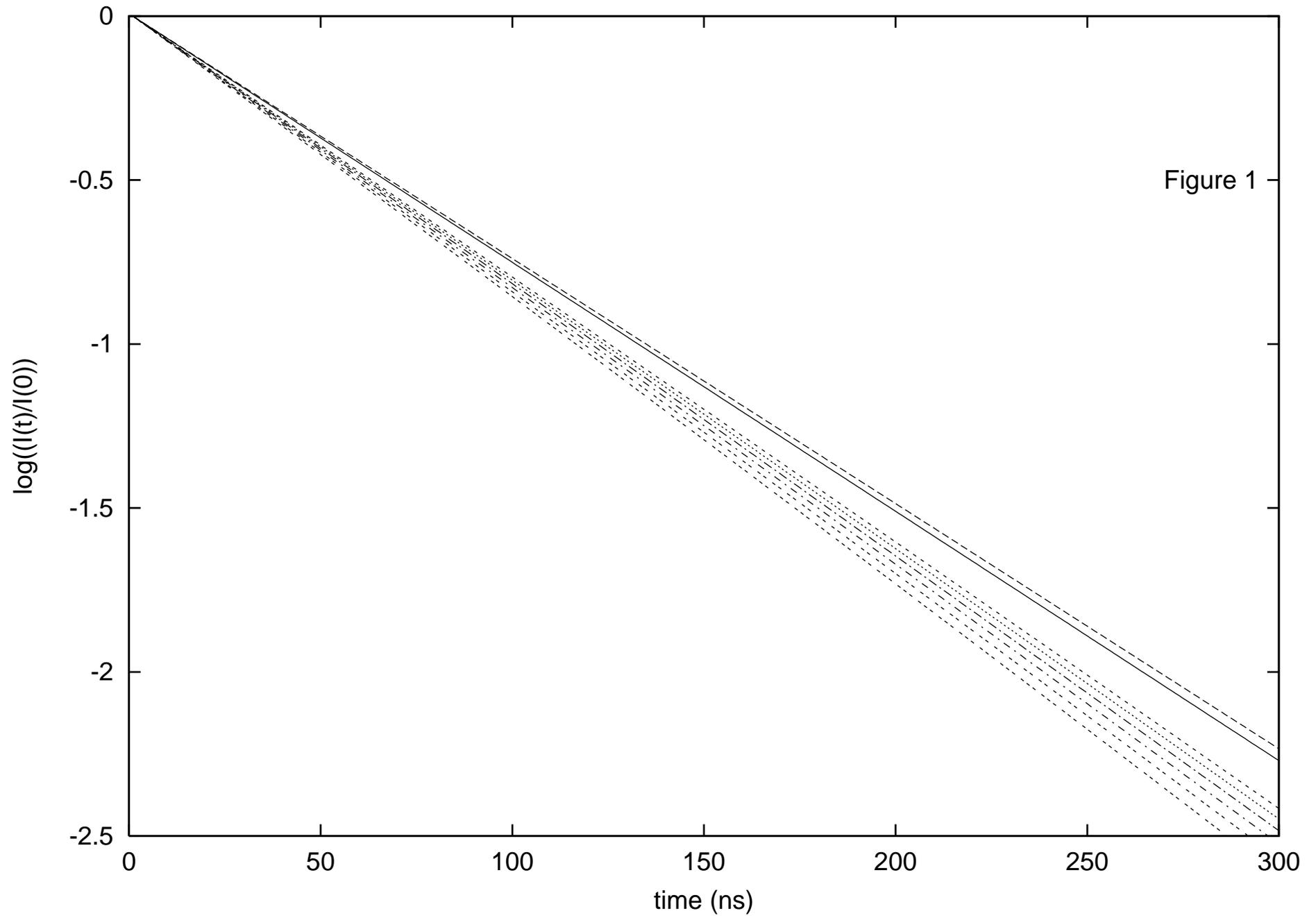
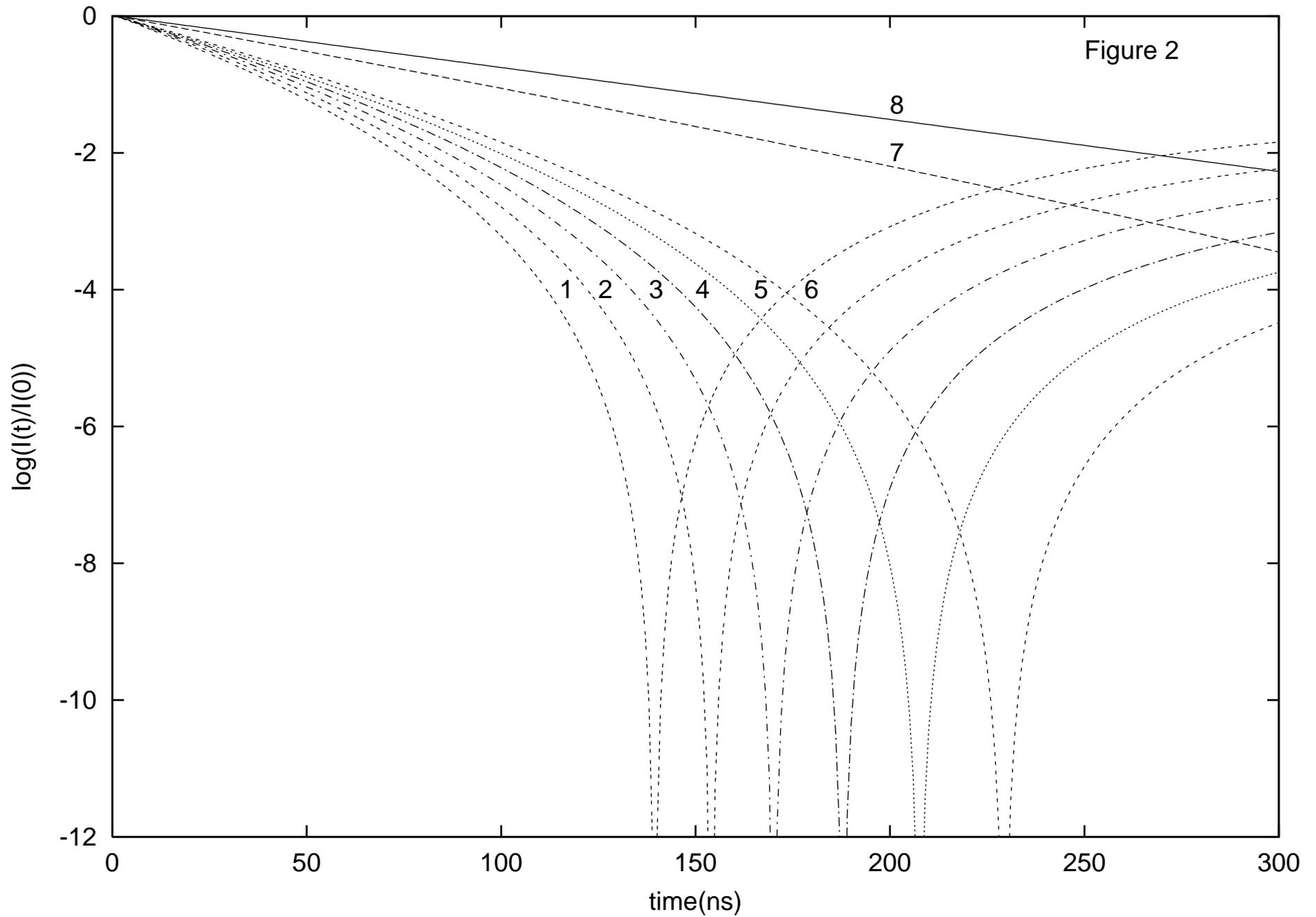
\end{document}